\documentclass[aps,epsfig,twocolumn,showpacs,email,amssymb,nobibnotes]{revtex4}
\usepackage{graphicx,amsmath,amssymb,psfrag,epsfig,multirow,slashbox}
\begin{document}
\title{Phase lag in epidemics on a network of cities}

\author{G. Rozhnova$^{1,2}$, A. Nunes$^1$ and A. J.~McKane$^{1,2}$}

\affiliation{$^1$Centro de F{\'\i}sica da Mat{\'e}ria Condensada and 
Departamento de F{\'\i}sica, Faculdade de Ci{\^e}ncias da Universidade de 
Lisboa, P-1649-003 Lisboa Codex, Portugal \\
$^2$Theoretical Physics Division, School of Physics and Astronomy,
University of Manchester, Manchester M13 9PL, United Kingdom
}

\begin{abstract}
We study the synchronisation and phase-lag of fluctuations in the number of 
infected individuals in a network of cities between which individuals commute. 
The frequency and amplitude of these oscillations is known to be very well 
captured by the van Kampen system-size expansion, and we use this approximation
to compute the complex coherence function that describes their correlation. We 
find that, if the infection rate differs from city to city and the coupling 
between them is not too strong, these oscillations are synchronised with a well
defined phase lag between cities. The analytic description of the effect is 
shown to be in good agreement with the results of stochastic simulations for 
realistic population sizes.
\end{abstract}
\maketitle

\section{Introduction}
\label{intro}

The theory of the frequency and amplitude of stochastic oscillations in models
of epidemics of childhood diseases has been extensively developed over the last
few years~\cite{alonso,STdG07,PhysRevE.79.041922,PhysRevE.80.051915,parisi,seasonalRN,epidforcingmckane,stagedmodelAndrew,kuske,ghose},  
but the question of the synchrony of these oscillations has received 
comparatively little attention. This is despite its undoubted importance; 
whether the oscillations in different locations are in phase or out of phase 
with each other will clearly have consequences for the duration of an epidemic 
and for the persistence of a disease. The particular case of in-phase and 
antiphase locking of epidemics in coupled cities has been described and 
interpreted in the literature~\cite{REG99,hestone}, but only for models that 
exhibit periodic oscillations in the infinite population limit. Similarly,
synchronisation in models of coupled oscillators has been extensively studied~\cite{synchbook},
including in particular demographic oscillators~\cite{galla08}, but not in the context of epidemics.

This subject has been so little investigated from a theoretical point of view,
that it is useful to recall a simpler, and more familiar example. It is
frequently argued in connection with predator-prey interactions that if
predator numbers happen to be high, prey numbers will subsequently fall due
to increased predation, and when prey numbers fall, predator numbers will
subsequently fall due to lack of food. This is actually an argument which 
relates to fluctuations in the number of predators and prey and would seem to
indicate oscillations, with predator and prey numbers being out of phase with 
each other. 

A similar argument, although not so well-known, can be applied to infected
and susceptible individuals in a city. In both cases, the frequency and 
amplitude of these oscillations is very well captured by the van Kampen 
system-size expansion~\cite{alonso,STdG07,PhysRevE.79.041922,PhysRevE.80.051915,parisi,seasonalRN,epidforcingmckane,stagedmodelAndrew,mckane_new2,van_kampen}, 
and it seems reasonable to suppose that the same approach should also be able 
to quantify the synchrony and phase-lag between oscillations of different 
kinds of individuals. In this paper we apply this method to study possible 
synchronisation between fluctuations in the number of infected individuals in 
a network of cities between which individuals commute. 

A previous study~\cite{epidcities} has shown that if the parameter setting the 
infection rate, denoted by $\beta$, is the same for all cities in the network, 
then the stochastic dynamics takes a remarkably simple form. This has been 
known for some time in the case of the deterministic dynamics: a rather general 
theorem tells us that in this case the fraction of susceptible, infected and 
recovered individuals is the same in all cities~\cite{guo-etal2008}. In 
Ref.~\cite{epidcities}, it was shown that the stochastic oscillations in this 
situation are also quite simple, having only one frequency, which is 
independent of the city, even if the sizes of the cities are different. As we 
will show later, these oscillations are synchronised between cities, but with 
no phase lag, in agreement with the findings reported in the literature for 
stochastic simulations of infection dynamics on coupled population 
patches~\cite{lloyd_may,REG99}. Exploring spatial heterogeneity as a means to 
overcome some of the shortcomings of the standard description of epidemics, 
namely in the predicted frequency of disease fade-outs, was one of the goals of
these studies. In this sense, the trivial phase relation between cities is a 
negative result.

However, it was assumed in previous work that the infection rate $\beta$ 
is the same in all cities. If $\beta$ is taken to be different in different 
cities, the symmetry of the fixed point is broken, which introduces the 
possibility of more complex oscillations. There are other ways of achieving
this, but within the framework in which we will work, this is a natural way of
introducing the symmetry breaking. We shall see that, under these conditions, 
the numbers of infected individuals fluctuate with a phase lag between cities,
provided that the coupling is not too strong. 

The paper is structured as follows. In Sec.~\ref{model} we describe the 
stochastic model, including a microscopic interpretation of the parameters in 
terms of the mobility patterns of the populations and the characteristics of 
the disease. The deterministic equations in the infinite population limit are 
derived in Sec.~\ref{dyn}, as are the Langevin equations for the fluctuations
around the non-trivial deterministic equilibrium in the linear noise 
approximation, found using van Kampen's system-size expansion. In 
Sec.~\ref{analyse} we introduce the complex coherence function that measures 
the cross-correlations between different cities and/or types of individuals, 
and we compute it for susceptibility and infection in the simple case of one 
city as an illustration of the method. The correlation between infection in 
two cities is studied in Sec.~\ref{2cities} and it is shown that for moderate 
coupling and different $\beta$ the fluctuations have a characteristic frequency
range and phase relation. This is extended to the case of three cities in 
Sec.~\ref{3cities}. We conclude in Sec.~\ref{final}.

\section{Model}
\label{model}
The model consists of $n$ cities, labeled $j=1,\ldots,n$ containing 
individuals who are infected, susceptible or recovered from a disease. As
is common, for mathematical convenience the total number of individuals in 
a particular city, $N_j$ for city $j$, is fixed. This means that the number 
of recovered individuals in each city may be expressed in terms of the number 
of infected and the number of susceptible individuals: $R_{j}=N_{j}-S_{j}-I_{j}$,
$j=1,\ldots,n$. This reduces the number of degrees of freedom from three to
two per city, considerably simplifying the analysis. It also means that 
deaths and births are coupled, with the random death of an individual giving
rise to the birth of a susceptible individual, and so providing a fresh victim 
for the disease. 

We will have in mind a particular type of interchange of individuals between
cities, but we will see that the construction which comes from this is 
sufficiently general to encompass a large variety of situations. A concrete
example of what we have in mind is a set of residential areas and one or more 
urban areas; there would typically be a significant fraction of commuters 
from the former to the latter, and fewer from the latter to the former. The
model will thus allow for a fraction of the individuals from city $j$ to
commute to city $k$, denoted by $f_{kj}$, with the number of non-commuters 
(residents) in city $j$ being $1-f_{j} \equiv 1-\sum_{k\neq j}f_{kj}$. The 
model will differ from that studied in Ref.~\cite{epidcities} in that we will 
take into account the different nature of the city (for instance, residential 
or urban) by allowing the parameter describing the rate of infection, $\beta$, 
to vary from city to city. 

If we denote the state of the system of $n$ cities at a given time by
$\sigma \equiv \{S_1,I_1,\ldots,S_n,I_n\}$, then the recovery of an individual
in city $j$ will occur at a rate $\gamma I_j$ and cause a transition to the
new state $\sigma' = \{S_1,I_1,\ldots,S_j,I_{j}-1,\ldots,S_n,I_n\}$. We will 
write the transition rate for this case as 
\begin{equation}
T(S_j,I_{j}-1|S_j,I_j)=\gamma I_j, 
\label{recovery}
\end{equation}
with the initial state on the right and the final state on the left. It should 
be noted that we have only listed only the variables in city $j$ in order to 
lighten the notation. In a similar fashion, the death of an infected or
recovered individual, and the birth of a susceptible, in city $j$ are given by
\begin{eqnarray}
T(S_{j}+1,I_{j}-1|S_j,I_j) &=& \mu I_j , \nonumber \\
T(S_{j}+1,I_j|S_j,I_j) &=& \mu (N_{j}-S_{j}-I_{j}), 
\label{death_birth}
\end{eqnarray}
respectively. 
  
The network structure manifests itself when constructing the transition rates
induced by infections. The rate of infections involving susceptibles from
city $j$ and infectives from city $k$ will be proportional to $S_{j}I_{k}$.
There will be five types of term. In two of them $k=j$. These are when the
infective residents in city $j$ infect susceptible residents in
city $j$ and when infective commuters from city $j$ infect susceptible 
commuters from city $j$ in city $\ell$ ($\ell \neq j$). The rates for 
these are respectively $\beta_j (1-f_j)^2 S_j I_j/M_j$ and
$\beta_{\ell} f^2_{\ell j} S_{j} I_j/M_{\ell}$, where $\beta_j$ is the parameter 
which characterises the infection rate in city $j$ and where $M_j$ is the 
number of individuals in city $j$:
\begin{equation} 
M_j = \left( 1 - f_j \right)N_j + \sum_{m\neq j} f_{jm}N_{m}.
\label{M}
\end{equation}

The other three terms result when the susceptibles from city $j$ are infected 
by individuals from city $k$ where cities $j$ and $k$ are different. Then,
infective commuters can infect resident susceptibles at a rate
$\beta_j (1-f_j)f_{jk}S_jI_k/M_j$, infective residents can infect commuting 
susceptibles at a rate $\beta_k f_{kj} (1-f_k) S_jI_k/M_k$ and finally 
infective commuters from city $k$ infect susceptible commuters from city $j$ 
in city $\ell$ ($\ell \neq j,k$) at a rate 
$\beta_{\ell}f_{\ell j}f_{\ell k} S_jI_k/M_{\ell}$. These results allow us to 
write down the transition rate for infection as
\begin{equation}
T(S_{j}-1,I_{j}+1|S_j,I_j) = \sum^{n}_{k=1} \beta_{jk} \frac{S_j I_k}{N_k},
\label{infection}
\end{equation}
where
\begin{eqnarray}
\beta_{jj} &=& \frac{\beta_j \left( 1 - f_j \right)^2 N_j}{M_j}
+ \sum_{\ell \neq j} \frac{\beta_{\ell}f_{\ell j}^2 N_j}{M_\ell}, \ \ j=1,\ldots,n,
\nonumber \\
\beta_{jk} &=& \frac{\beta_j \left( 1 - f_j \right)f_{jk} N_k}{M_j} 
+ \frac{\beta_k f_{kj}\left( 1 - f_k \right) N_k}{M_k} \nonumber \\
&+& \sum_{\ell\neq j,k} \frac{\beta_{\ell} f_{\ell j} f_{\ell k} N_k}{M_\ell}, \ \
j,k=1,\ldots,n; j \neq k.
\label{beta_jk}
\end{eqnarray}

Although we have a particular picture of how individuals move between cites,
and of the assignment of infection rates to the cities themselves, the final
form of the transition rate (\ref{infection}) is quite general. Our results 
will therefore apply to a wide range of the possible types of interchanges of 
individuals and choice of infection parameters. Unlike the case where the 
$\beta_j$ were all equal~\cite{epidcities}, the $\beta_{jk}$ have no relations 
between them, and so in general are independent. Therefore they can be chosen 
from a consideration of Eq.~(\ref{beta_jk}) or simply externally imposed.

\section{Dynamics}
\label{dyn}
The model defined in Section~\ref{model} is stochastic and Markovian, since
the transition rates (\ref{recovery}), (\ref{death_birth}) and (\ref{infection})
do not depend on past states of the system. This means that adopting a 
continuous time description, the time evolution of the system may be obtained
from a master equation for $P(\sigma,t)$, the probability distribution for 
finding the system in state $\sigma$ at time $t$~\cite{van_kampen,Risken,Gardiner} 
\begin{equation}
\frac{dP(\sigma,t)}{dt} = \sum_{\sigma' \neq \sigma}
\left[ T(\sigma | \sigma')P(\sigma',t) - 
T(\sigma' | \sigma)P(\sigma,t) \right],
\label{master}
\end{equation}
where the $T(\sigma | \sigma')$ are each of the rates (\ref{recovery}),
(\ref{death_birth}) and (\ref{infection}) taken in turn~\cite{van_kampen,Risken,Gardiner}.

The full master equation (\ref{master}) cannot be solved exactly, but the
aspects that concern us here can be analysed in a remarkably precise way using 
the system-size expansion of van Kampen~\cite{van_kampen}. We have carried out 
this calculation on a simpler version of this model --- when all the
$\beta_j$ were equal --- elsewhere~\cite{epidcities}, and while the predictions
are different, the method of carrying out the expansion is very similar. A 
comparison of the transition rates shows that the results for the model we
are investigating here may be obtained from those in Ref.~\cite{epidcities} by 
substituting $\beta c_{jk}$ by $\beta_{jk}$. We will therefore only briefly 
indicate the steps in the analysis, and refer the reader to 
Ref.~\cite{epidcities} for more details.

The method begins by making the following ansatz \cite{van_kampen}:
\begin{equation}
S_j = N_js_j+N_j^{1/2}x_j,\ \ I_j=N_ji_j+N_j^{1/2}y_j,
\label{ansatz}
\end{equation}
where $j=1,\ldots,n$. Here $s_j = \lim_{N_j \to \infty} S_j/N_j$ and 
$i_j = \lim_{N_j \to \infty} I_j/N_j$ are the fraction of individuals from city
$j$ which are respectively susceptible and infected in the deterministic 
limit. Therefore the variables are broken down into a sum of deterministic
terms, $s_j$ and $i_j$, and the stochastic deviations from these, $x_j$ and 
$y_j$. The deterministic terms satisfy the ordinary differential equations
\begin{eqnarray}
\dfrac{ds_j}{dt}&=& - \sum^{n}_{k=1} \beta_{jk} s_j i_k + 
\mu \left( 1 - s_j \right), \nonumber \\
\dfrac{di_j}{dt}&=& \sum^{n}_{k=1} \beta_{jk} s_j i_k -
\left( \gamma + \mu \right) i_j,
\label{deter}
\end{eqnarray}
where $j=1,\ldots,n$. 

The stochastic fluctuations, $x_j$ and $y_k$, which describe the linear 
fluctuations around trajectories of the deterministic set of equations 
(\ref{deter}), are found to obey a set of linear stochastic differential 
equations. For convenience we introduce the vector of these fluctuations 
$\mathbf{z}=(x_1,\ldots,x_n,y_1,\ldots,y_n)$ and indices $J,K=1,\ldots,2n$,
Then these stochastic differential equations have the 
form~\cite{van_kampen,Risken,Gardiner}
\begin{equation}
\frac{d z_J}{dt} = \sum_{K=1}^{2n} A_{JK} z_K + \eta_J(t), \ \ J=1,\ldots,2n,
\label{Langevin}
\end{equation}
where $\eta_J(t)$ are Gaussian noise terms with zero mean which satisfy 
$\langle\eta_J(t)\eta_K(t')\rangle = B_{JK}\delta(t-t')$. The two $2n \times 2n$
matrices $A$ and $B$ are obtained from the expansion procedure. In fact, to
determine the matrix $A$ it is not necessary to carry out the full 
system-size expansion, since it is related to the Jacobian, ${\cal J}$, found 
from linear stability analysis about a fixed point of Eq.~(\ref{deter}). The
precise relationship is ${\cal J}=S^{-1}AS$, where 
$S={\rm diag}(\sqrt{N_1},\ldots,\sqrt{N_n})$~\cite{epidcities}. The explicit 
forms for ${\cal J}$ and $B$ are most easily given in terms of the four 
$n \times n$ submatrices:
\begin{equation}
{\cal J} =\left[\begin{array}{c|c} 
{\cal J}^{(1)} & {\cal J}^{(2)} \\ \hline 
{\cal J}^{(3)} & {\cal J}^{(4)} 
 \end{array}\right],
\label{block}
\end{equation}
and similarly for $B$. The elements of these submatrices are found to be
\begin{eqnarray}
{\cal J}^{(1)}_{jk}&=&-\mu\delta_{jk}-\delta_{jk}\sum_{\ell=1}^n 
\beta_{j\ell}i_{\ell},
\nonumber \\
{\cal J}^{(2)}_{jk}&=&-s_j\beta_{jk},
\nonumber \\
{\cal J}^{(3)}_{jk}&=&\delta_{jk}\sum_{\ell=1}^n \beta_{j\ell}i_{\ell},
\nonumber \\
{\cal J}^{(4)}_{jk}&=&-(\mu+\gamma)\delta_{jk}+ s_j\beta_{jk},
\label{J_entries}
\end{eqnarray}
and
\begin{eqnarray}
B^{(1)}_{jk} &=& \mu\delta_{jk}\left( 1 - s_j \right) + \delta_{jk} 
\sum_{\ell=1}^n s_j \beta_{j\ell} i_{\ell}, \nonumber \\
B^{(2)}_{jk}=B^{(3)}_{jk}&=& -\mu\delta_{jk} i_{j} -\delta_{jk} 
\sum_{\ell = 1}^n s_j \beta_{j\ell} i_{\ell}, \nonumber \\
B^{(4)}_{jk}&=& \left( \gamma + \mu \right)\delta_{jk} i_j + \delta_{jk}
\sum_{\ell =1}^n s_j \beta_{j\ell} i_{\ell}.
\label{B_entries}
\end{eqnarray}
These matrices depend on the solutions $s_j$ and $i_j$ of Eq.~(\ref{deter}), 
and so on time, but since we will be interested in fluctuations about the 
stationary state, we use the fixed point values of $s_j$ and $i_j$, denoted 
as $s^*_j$ and $i^*_j$. The resulting matrices, ${\cal J}^*$ and $B^*$, are 
then time-independent.

\section{Analysis}
\label{analyse}
Adding the two sets of equations in (\ref{deter}), we immediately see that
the fixed points satisfy
\begin{equation}
\left( \gamma + \mu \right)i^{*}_j = \mu\left( 1 - s^{*}_j \right), \ \ 
j=1,\ldots,n.
\label{first_FP_eqn}
\end{equation}
Using this equation to eliminate the $i^{*}$, one finds that
\begin{equation}
s^{*}_j \left[ \left( \gamma + \mu \right) +
\sum^n_{k=1}\beta_{jk} \left( 1 - s^{*}_k \right) \right] = 
\left( \gamma + \mu \right), \ \  j=1,\ldots,n.
\label{second_FP_eqn}
\end{equation}

We will assume that the matrix of the coupling coefficients $\beta_{jk}$ is 
irreducible, which means that any two cities have a direct or indirect 
interaction. Otherwise the $n$ cities may be split into non-interacting 
subsets, and several equilibria may be found by combining disease extinction 
in some subsets with non-trivial equilibrium in other subsets. The disease
extinction fixed point is simple to find. If we assume that any one of the  
$i^{*}$ is zero, for instance $i^{*}_{\ell} = 0$, then from 
Eq.~(\ref{first_FP_eqn}) $s^{*}_{\ell} = 1$. From Eq.~(\ref{deter}) we see 
immediately that $\sum^{n}_{k=1} \beta_{\ell k} i^{*}_k = 0$. Since the matrix
of couplings is irreducible, and from Eq.~(\ref{beta_jk}) the entries are
non-negative, it follows that $i^{*}_k = 0$ for all $k$.

Although it is difficult, and in most cases impossible, to determine any
non-trivial fixed points analytically, a theorem~\cite{guo-etal2008} tells us 
that a unique stable fixed point exists and is stable. The linear stochastic 
deviations from this fixed point satisfy Eq.~(\ref{Langevin}), with $A$ and 
$B$ being evaluated at the fixed point. To find the dominant frequencies of 
the stochastic oscillations, and as we shall see also to study synchronisation
and phase-lag, we Fourier transform Eq.~(\ref{Langevin}) to obtain
\begin{equation}
\sum^{2n}_{K=1} \left( -i\omega \delta_{JK} - A_{JK} \right) \tilde{z}_{K} 
(\omega) = \tilde{\eta}_{J}(\omega), \ \ J=1,\ldots,2n,
\label{FT_Lang}
\end{equation}
where the $\tilde{f}$ denotes the Fourier transform of the function $f$.
Defining the matrix $-i\omega \delta_{JK} - A_{JK}$ to be $\Phi_{JK}(\omega)$,
the solution to Eq.~(\ref{FT_Lang}) is
\begin{equation}
\tilde{z}_{J}(\omega) = \sum^{2n}_{K=1} \Phi^{-1}_{JK}(\omega) 
\tilde{\eta}_{K}(\omega).
\label{solution}
\end{equation}

We now introduce the matrix
\begin{eqnarray}
P_{JK}(\omega) &\equiv& \langle \tilde{z}_J(\omega) 
\tilde{z}^*_K(\omega) \rangle \nonumber \\
&=& \sum^{2n}_{L=1} \sum^{2n}_{M=1} \Phi^{-1}_{JL}(\omega)B_{LM}
\left( \Phi^{\dag} \right)^{-1}_{MK}(\omega),
\label{defn}
\end{eqnarray}
where $*$ now denotes complex conjugation.

In previous studies, where the focus was on finding the frequencies and 
amplitudes of the stochastic oscillations~\cite{alonso,STdG07,PhysRevE.79.041922,PhysRevE.80.051915,parisi,seasonalRN,epidforcingmckane,stagedmodelAndrew,mckane_new2}, 
only the power spectrum (when $J=K$) was analysed. Here we will also 
be interested in the cross-correlations between infection in two different 
cities, and so will also wish to calculate the cross-spectrum (when 
$J \neq K$). It is frequently convenient to normalise this by the relevant 
power-spectrum, and instead work with the complex coherence function (CCF) 
defined by~\cite{neiman,Marple,Stoica}
\begin{equation}
C_{JK}(\omega) \equiv \frac{P_{JK}(\omega)}{\sqrt{P_{JJ}(\omega)P_{KK}(\omega)}}.
\label{CCF}
\end{equation}

The CCF will in general be complex for $J \neq K$, and so typically one 
calculates its magnitude and phase, that is, the coherence
\begin{equation}
\left| C_{JK}(\omega) \right| = 
\frac{\left| P_{JK}(\omega) \right| }{\sqrt{P_{JJ}(\omega)P_{KK}(\omega)}},
\label{coherence}
\end{equation}
and the phase spectrum
\begin{equation}
\phi_{JK}(\omega) \equiv 
\tan^{-1}\left[ \frac{{\rm Im}\left( C_{JK}(\omega) \right)}
{{\rm Re}\left( C_{JK}(\omega) \right)} \right] =
\tan^{-1}\left[ \frac{{\rm Im}\left( P_{JK}(\omega) \right)}
{{\rm Re}\left( P_{JK}(\omega) \right)} \right].
\label{phase_spect}
\end{equation}

\begin{figure}
\centering
\includegraphics[width=\columnwidth]{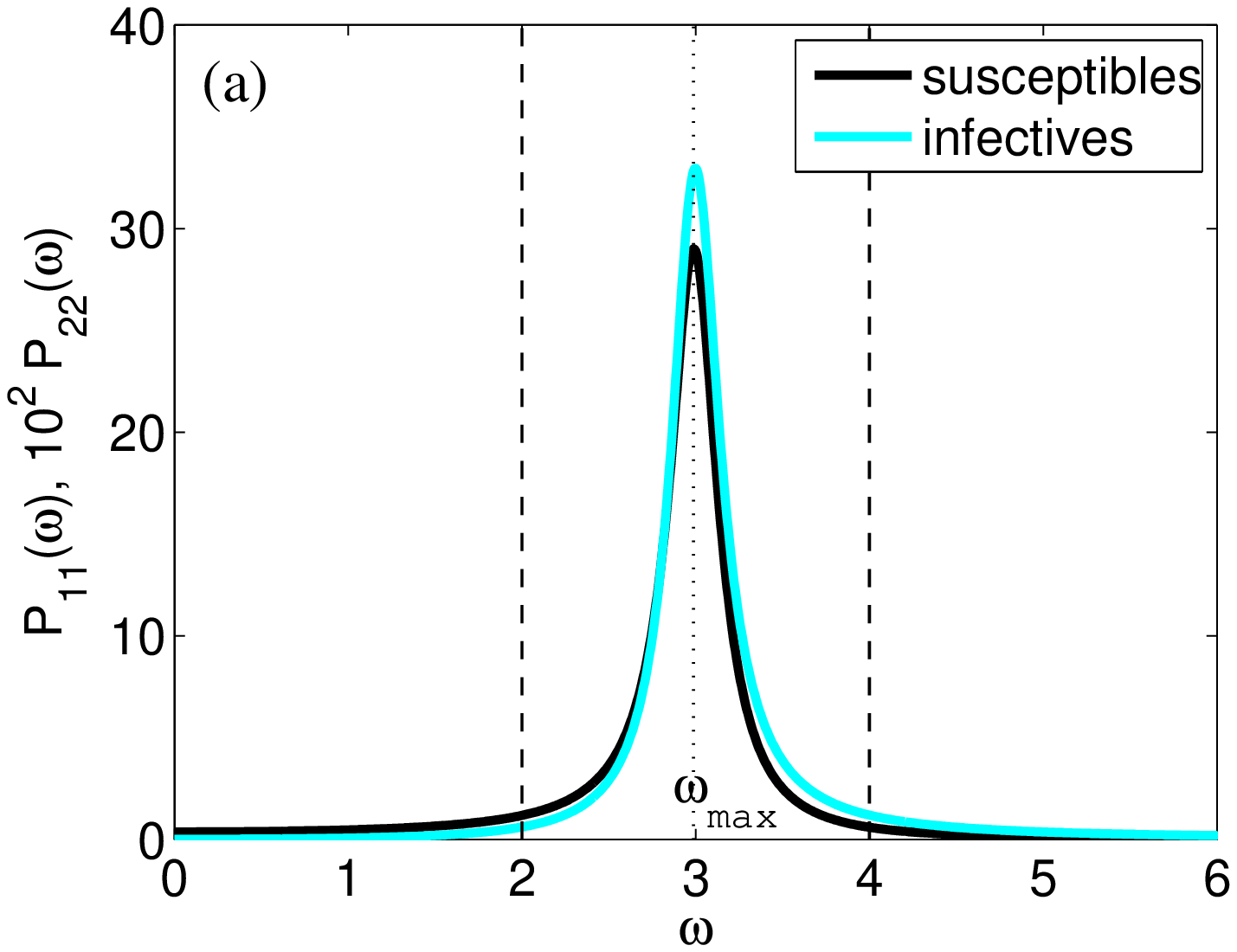}
\includegraphics[width=\columnwidth]{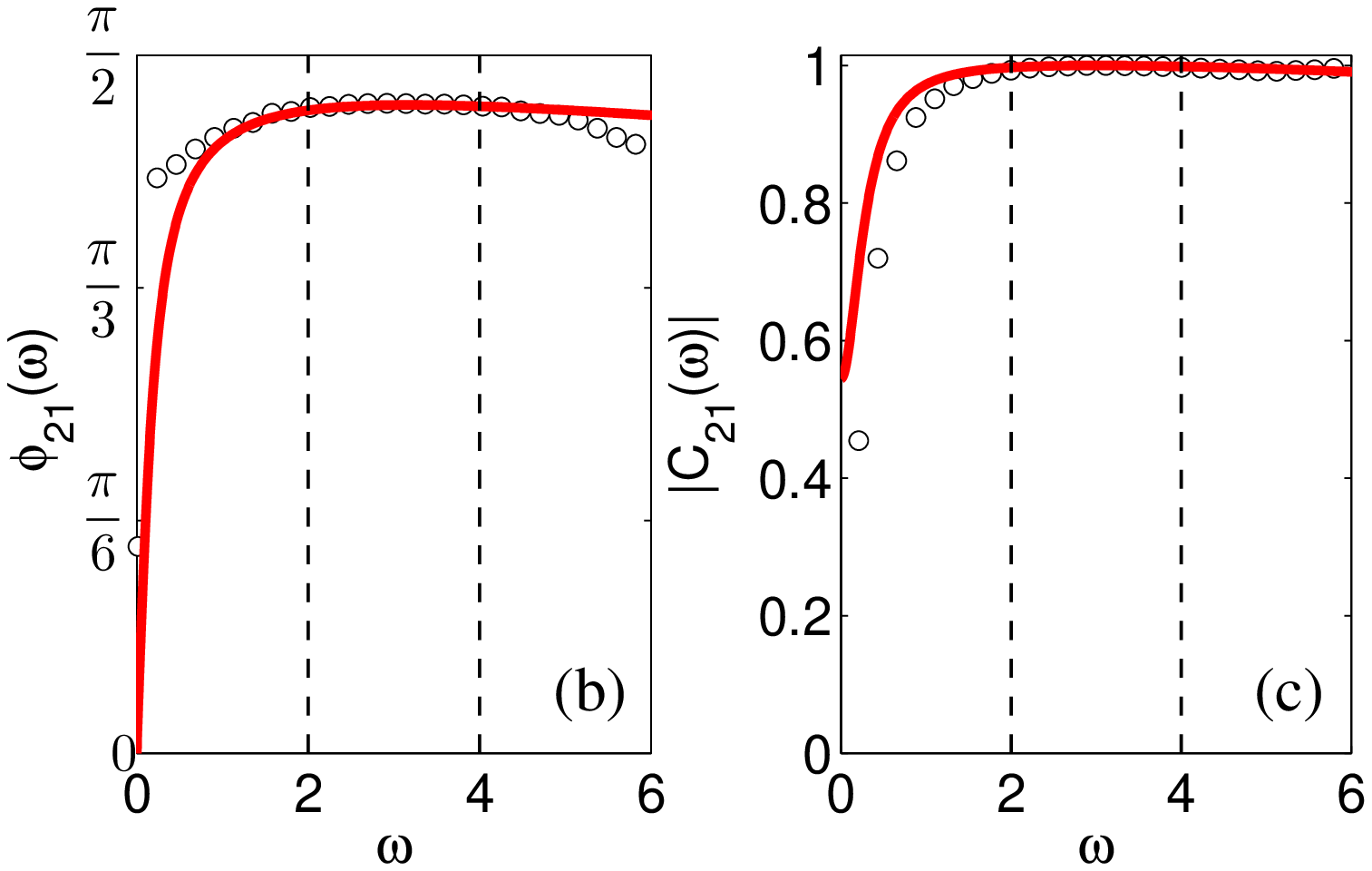}
{\caption{(Color online) Synchronisation and phase-lag between susceptible and 
infected individuals in the one-city SIR model. (a) Analytical power spectra 
of the fluctuations of susceptibles (solid black line), $P_{11}(\omega)$, and of 
infectives [solid cyan (gray) line], $P_{22}(\omega)$, given by Eq.~(\ref{defn}). The
dotted vertical line indicates the frequency $\omega_{max}$ at which these 
spectra attain maxima. (b) Phase spectrum, $\phi_{21}(\omega)$, and (c) 
coherence, $\left| C_{21}(\omega) \right|$, for the fluctuations of 
infectives and susceptibles. The solid lines are the analytical results given 
by Eq.~(\ref{PS_simple}) and Eq.~(\ref{coherence_simple}). The open circles are
the results for the same quantities obtained from simulations. Parameters: 
$N=10^6$, $\beta=17(\gamma+\mu)$, $\mu=1/50$ 1/y and 
$\gamma=365/13$ 1/y. In all panels the dashed vertical lines bound
the frequency range where the stochastic amplification is significant.}}
\label{fig1}
\end{figure}

As an example of using the CCF to understand synchronisation and phase-lag in 
systems with sustained stochastic cycles, we apply it to the SIR model in a 
single city. We could also apply it to the predator-prey system discussed 
earlier, but we already have the required equations for the one-city SIR model 
in this paper: Eqs.~(\ref{deter})-(\ref{second_FP_eqn}), with the indices 
omitted and with $\beta_{jk}$ replaced by $\beta$. It should be emphasised that
in this example we are looking at the synchronisation and phase-lag between 
susceptible and infected individuals, whereas in the actual application of the 
formalism to $n$ cities, the interest is in possible synchronisation and 
phase-lag between infected individuals in city $j$ and infected individuals in 
city $k$.

For models such as the one-city SIR and the predator-prey model, 
Eq.~(\ref{coherence}) becomes
\begin{equation}
\left| C_{21}(\omega) \right| =
\sqrt{\frac{g_4\omega^4 +g_2 \omega^2+g_0}{h_4\omega^4 +h_2 \omega^2+h_0} },
\label{coherence_simple}
\end{equation}
where the coefficients of the polynomials are sums of products of the $A_{JK}$ 
and $B_{JK}$, $J,K=1,2$, which can be found from
Eqs.~(\ref{FT_Lang})-(\ref{coherence}), and Eq.~(\ref{phase_spect}) becomes 
simply
\begin{equation}
\phi_{21}(\omega)=\tan^{-1}\left[ \frac{k_1 \omega}{k_0 + k_2\omega^2}\right],
\label{PS_simple}
\end{equation}
where again $k_{0}, k_{1}$ and $k_{2}$ are sums of products of the $A_{JK}$ and 
$B_{JK}$, $J,K=1,2$.

\begin{figure}
\centering
\includegraphics[width=\columnwidth]{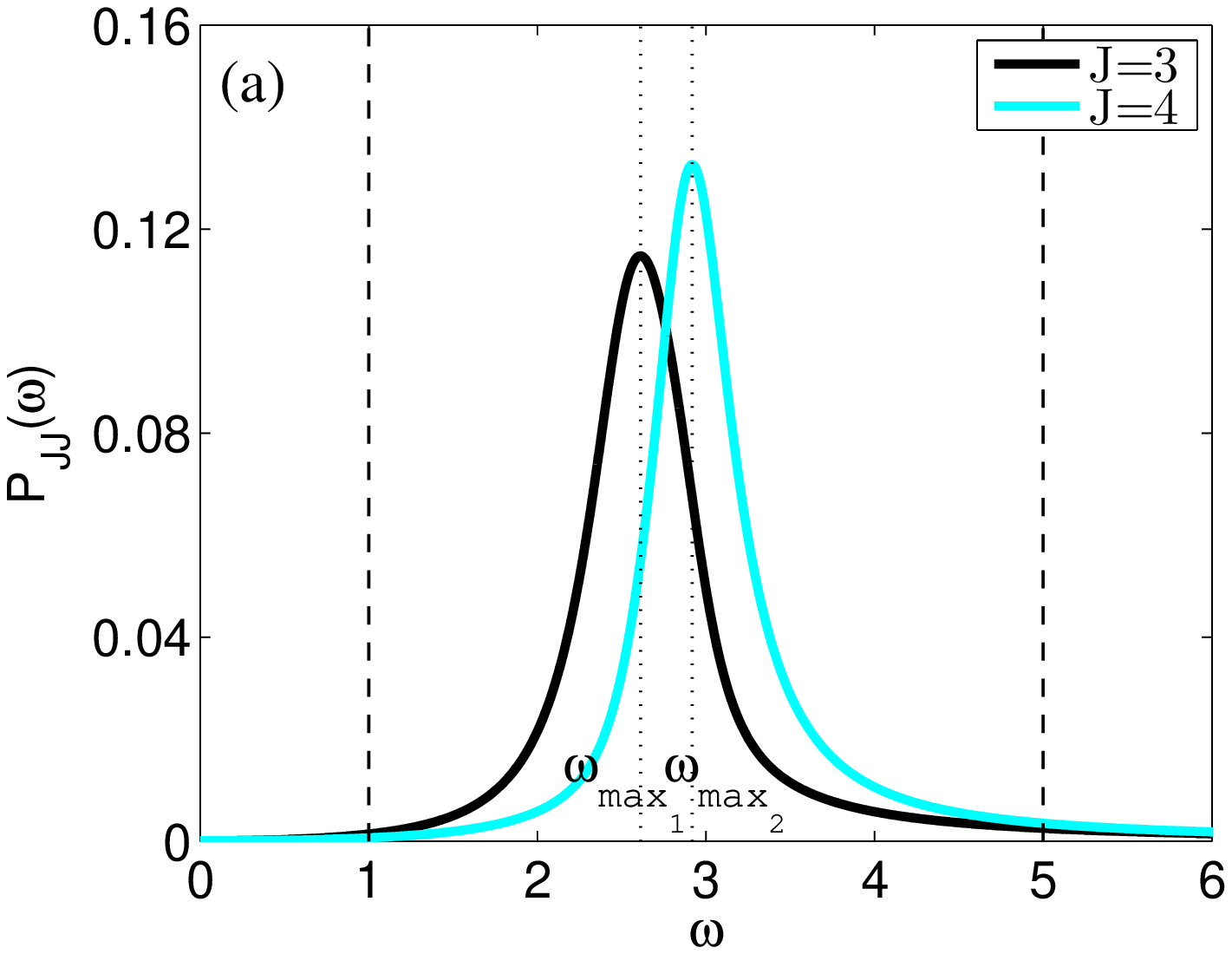}
\includegraphics[width=\columnwidth]{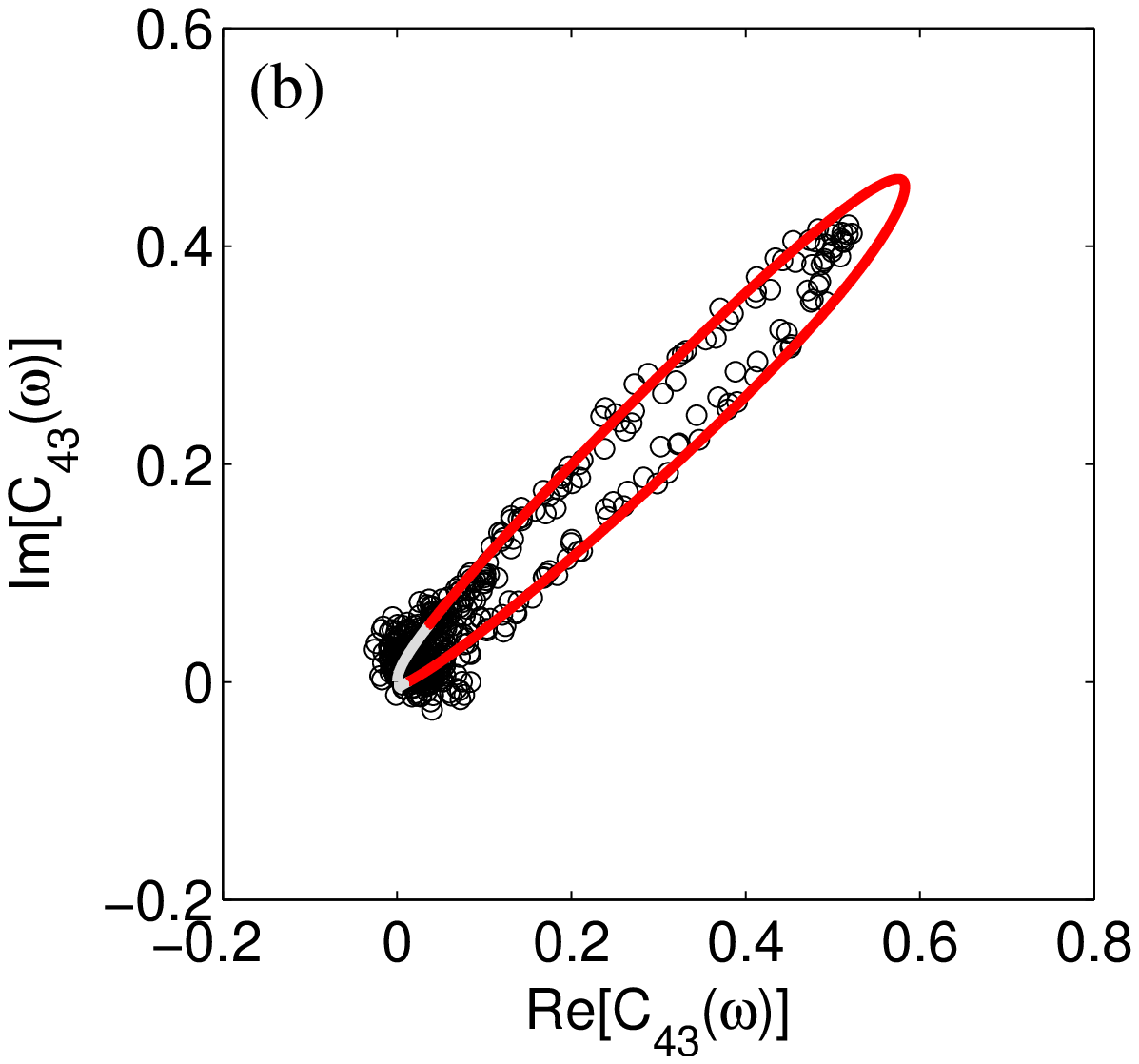}
{\caption{(Color online) Synchronisation and phase-lag between infected 
individuals in city 1 and infected individuals in city 2 in the two-city SIR 
model. (a) Power spectra of the fluctuations of infectives in city 1 (solid black line), 
$P_{33}(\omega)$, and of infectives in city 2 [solid cyan (gray) line], $P_{44}(\omega)$, given by 
Eq.~(\ref{defn}). The dotted vertical lines indicate the frequencies
$\omega_{max_1}$ and $\omega_{max_2}$ at which 
these spectra attain maxima. The dashed vertical lines bound the frequency 
range where the stochastic amplification is significant. (b) Parametric 
plot of the CCF, $C_{43}(\omega)$, in the complex plane. The solid line is the 
analytical result given by Eq.~(\ref{CCF}) [the parts of the curve shown in 
red (dark gray) and in light gray correspond to $1\leq\omega\leq5$ and 
$\{ 0\leq\omega<1 \}\cup\{ 5<\omega\leq100\}$, respectively] and the open 
circles are the result for the same quantity obtained from simulations for 
$0\leq\omega\leq 10$. Parameters: $\beta_1=12(\gamma+\mu)$, 
$\beta_2=17(\gamma+\mu)$, $N_1=N_2=10^6$, $f_{12}=0.001$, $f_{21}=0.01$, 
$\mu=1/50$ 1/y and $\gamma=365/13$ 1/y.}}
\label{fig2}
\end{figure}

The power spectra of the fluctuations of susceptible and infected individuals 
computed from Eqs.~(\ref{defn}), $P_{11} (\omega )$ and $P_{22} (\omega )$, are 
shown in Fig.~1 (a). Stochastic amplification gives rise to 
significant fluctuations in a well-defined frequency range centered at the 
frequency $\omega_{max}$ where both $P_{11} (\omega )$ and $P_{22} (\omega )$ 
peak. The coherence, $\left|C_{21}(\omega)\right|$, and phase spectrum, 
$\phi_{21}(\omega)$, given by Eq.~(\ref{coherence_simple}) and 
Eq.~(\ref{PS_simple}) are shown in Fig.~1 (c) and (b), respectively, 
together with the results for the same quantities obtained from numerical 
simulations. There is very good agreement between the analytic approximation 
and the simulations in the frequency range where the fluctuations have 
significant amplitude. Outside of this range, the magnitudes of the power 
spectra are so low that errors become significant, leading to agreement which
is not as good. However within this frequency range the fluctuations of the 
two kinds of individuals are strongly correlated and have a well-defined phase 
lag.

The numerical results presented here and in the following sections were 
obtained from long numerical simulations based on the Gillespie 
algorithm~\cite{gillespie}. Each run started from a random initial condition,
with the vector of the fluctuations of susceptibles and infectives in $n$ 
cities, $\mathbf{z}=(x_1,\ldots,x_n,y_1,\ldots,y_n)$, computed from simulated 
time series according to Eq.~(\ref{ansatz}), and recorded at equal time 
intervals. From each simulation run the power and cross spectra for the 
fluctuations are computed as $\tilde{z}_J(\omega) \tilde{z}^*_K(\omega)$, 
where $J,K=1,\ldots,2n$ and where $*$ denotes complex conjugation, by using 
discrete Fourier transforms. The final spectra are averages of $10^3$ to 
$5\times10^3$ simulations. Having computed these, the numerical CCFs, coherence
and phase spectra can easily be computed from 
Eqs.~(\ref{CCF})-(\ref{phase_spect}). The values for the epidemiological and 
demographic parameters used in this paper are those relevant for measles. In 
all simulations, we take $\mu=1/50$ 1/y and $\gamma=365/13$ 
1/y~\cite{and_may,KR07,KeelingGrenfellMeasles}.

\section{Two cities}
\label{2cities}
We now turn to the description of the synchronisation and phase-lag between 
infected individuals in two different cities, using the quantities introduced 
in Sec.~\ref{analyse}. Plots of the power spectra for infectives, 
$P_{33} (\omega )$ and $P_{44} (\omega )$, are shown in Fig.~2 (a) 
for a certain choice of parameters (recall that $z_3$ and $z_4$ are the 
fluctuations of infection in cities $1$ and $2$ respectively). Different 
values have been taken for $\beta_1$ and $\beta_2$, to reflect 
different social contact patterns in the two cities. In contrast with the 
spectra for different types of individuals in the one-city case, 
$P_{33} (\omega )$ and $P_{44} (\omega )$ peak at different frequencies 
$\omega_{max_1}$ and $\omega_{max_2}$. The frequency range of interest is bounded
by the two dashed vertical lines in the figure --- outside this range, 
$P_{33}$ and $P_{44}$ have negligible amplitude. A parametric plot of the range 
of the coherence function $C_{43}(\omega)$ in the complex plane is shown in 
Fig.~2 (b), where the darker (red in the color version) portion of the 
full line corresponds to the frequency range highlighted in Fig.~2 (a). It 
can be seen that the fluctuations in the number of those infected in the two 
cities are correlated and synchronize with a well-defined phase relation in 
the whole frequency range where both their amplitudes are significant.

\begin{figure}
\centering
\includegraphics[width=\columnwidth]{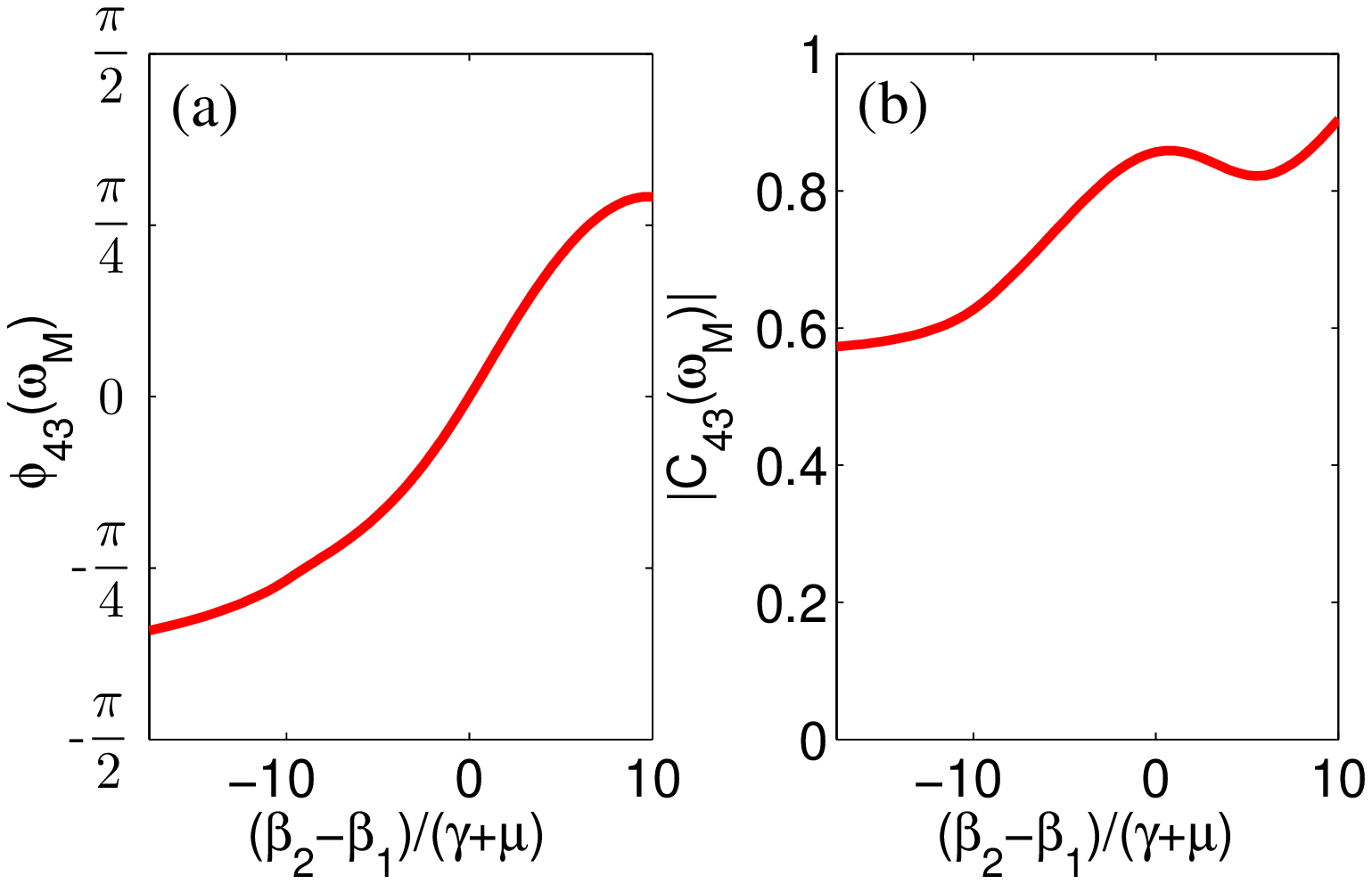}
\includegraphics[width=\columnwidth]{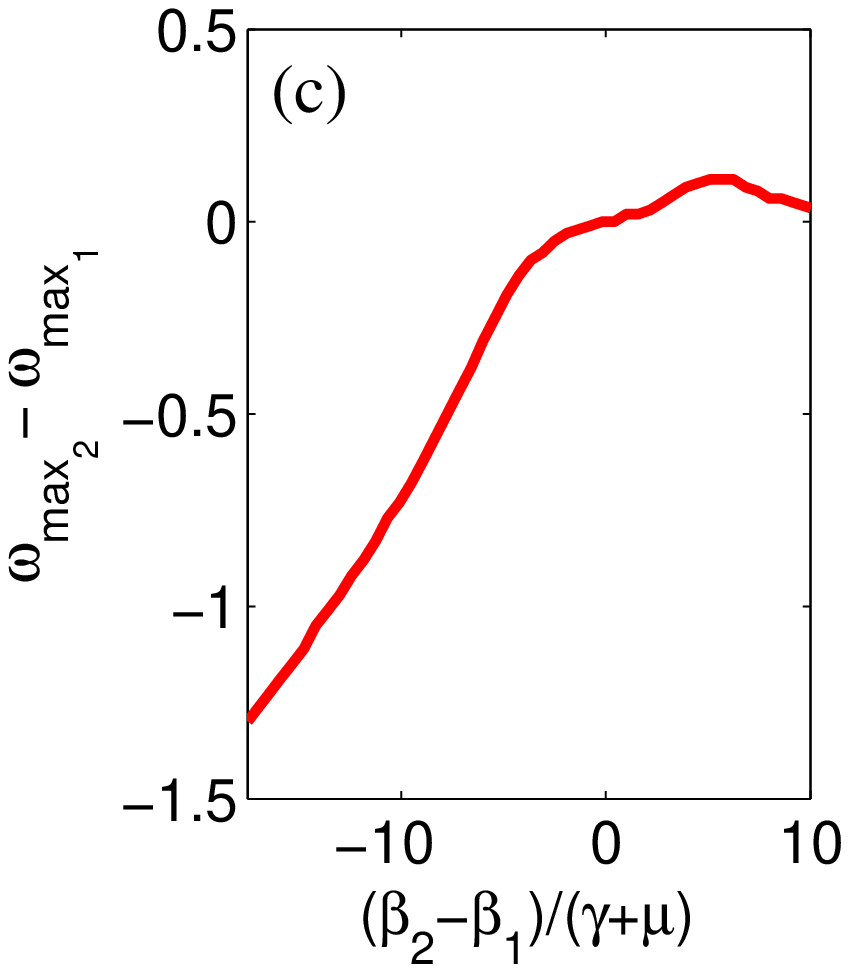}
{\caption{Dependence of the synchronisation and phase-lag between infected 
individuals in city 1 and infected individuals in city 2 on the infection 
rates in the two-city SIR model. (a) Phase, $\phi_{43}(\omega_M)$, 
and (b) coherence, $\left|C_{43}(\omega_M)\right|$, given by 
Eq.~(\ref{phase_spect}) and by Eq.~(\ref{coherence}) as a function of 
$(\beta_2-\beta_1)/(\gamma+\mu)$. (c) The difference between the peak 
frequencies $\omega_{max_1}$ of $P_{33}$ and $\omega_{max_2}$ of $P_{44}$ as a 
function of $(\beta_2-\beta_1)/(\gamma+\mu)$. To obtain the plots 
$\beta_2/(\gamma+\mu)$ was fixed at $15$ and $\beta_1/(\gamma+\mu)$ was varied 
between $5$ and $32.5$. Parameters: $N_1=10^6$, $N_2=2\times10^6$, 
$f_{12}=0.005$, $f_{21}=0.01$, $\mu=1/50$ 1/y and $\gamma=365/13$ 1/y.}}
\label{fig3}
\end{figure}

\begin{figure}
\centering
\includegraphics[width=\columnwidth]{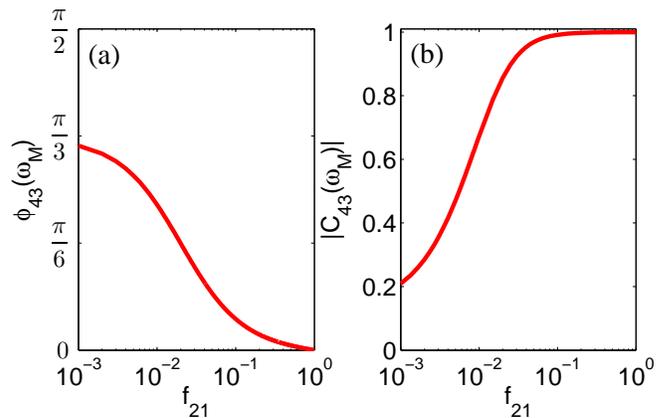}
{\caption{Dependence of the synchronisation and phase-lag between infected 
individuals in city 1 and infected individuals in city 2 on the fractions of 
commuters between the cities in the two-city SIR model. The description and 
coloring are as in Fig.~3. To obtain these log-linear plots, $f_{12}$ was fixed
at 0.001 and $f_{21}$ was varied between 0 and 1. Parameters: 
$\beta_1=12(\gamma+\mu)$, $\beta_2=17(\gamma+\mu)$, $N_1=10^6$, 
$N_2=2\times10^6$, $\mu=1/50$ 1/y and $\gamma=365/13$ 1/y.}}
\label{fig4}
\end{figure}

Also shown in Fig.~2 (b) are the results for $C_{43}(\omega)$ 
obtained from numerical simulations. There is a good agreement between 
simulations and analytic results within the frequency resolution limits of the 
former, showing that for the chosen system sizes the fluctuation 
cross-correlation behavior is well captured by the linear noise approximation.

In order to investigate the dependence of the coherence and phase spectrum
on the choice of parameters, we have computed the value $\omega_{M}$ for which
$\left|C_{43}(\omega)\right|$ attains its maximum, 
$\left| C_{43}(\omega_M)\right|$ and $\phi_{43}(\omega_M)$. 
A plot of  $\phi_{43}(\omega_M) $ and $\left|C_{43}(\omega_M)\right|$ as a 
function of $(\beta_2-\beta_1)/(\gamma+\mu)$ is shown in Fig.~3 (a) and (b). 
To obtain the plot only the infection rate $\beta_2$ was varied while the 
remaining parameters were kept fixed. With respect to the example of Fig.~2,
we have chosen one of the populations, $N_2$, to be twice as large and one of 
the coupling parameters, $f_{12}$, to be five times larger. It should be noted 
that when $\left(\beta_2-\beta_1\right)/(\gamma+\mu)$ is so large that the 
frequency ranges of the stochastic fluctuation peaks in each city no longer 
overlap, $\left | C_{43}(\omega) \right|$ becomes bimodal instead of unimodal 
and $\omega_{M}$ is no longer well defined. The difference between the peak 
frequencies $\omega_{max_1}$ of $P_{33}$ and $\omega_{max_2}$ of $P_{44}$ is 
shown in Fig.~3 (c). For values of 
$\left(\beta_2-\beta_1\right)/(\gamma+\mu)$ to the left of the represented 
range, i.e. for values smaller than $-17.5$, this difference becomes larger 
leading to a more complex synchronization pattern and a bimodal 
$\left | C_{43}(\omega) \right|$. 

We also see that, in this case, fluctuations with a non-trivial phase relation 
require the two cities to have different values of the infection rates 
$\beta_1$ and $\beta_2$. We have found this to be true in general. More 
precisely, the imaginary parts of the cross spectra of infection 
$P_{JK} (\omega)$ with $J,K=n+1,\ldots,2n$ and $J\neq K$, as given by 
Eq.~(\ref{defn}), are identically zero when the infection rates 
$\beta_i$, $i=1,\ldots,n$, are the same in all cities, independently of the 
choice of the population sizes and fractions of commuters. The proof of this 
result is given in the Appendix. This is compatible with either in-phase or 
antiphase synchronization. 

Another condition for $\phi_{43} (\omega _M) $ to be different from zero is 
that the coupling between the two cities is not too strong. A plot of 
$| C_{43}(\omega _M) |$ and $\phi_{43}(\omega _M) $ as a function of the 
coupling $f_{21}$ is shown in Fig.~4. The remaining parameters are as in Fig.~2,
except for $N_2$, which was taken twice as large. It can be seen that 
$\left| C_{43}(\omega _M) \right|$ increases monotonically with $f_{21}$, 
indicating an increase in cross correlation between infected individuals in 
the two cities as the coupling gets stronger, but $\phi_{43} (\omega _M) $ 
tends to zero with increasing $f_{21}$.

\section{Three cities}
\label{3cities}

\begin{figure}
\centering
\includegraphics[width=\columnwidth]{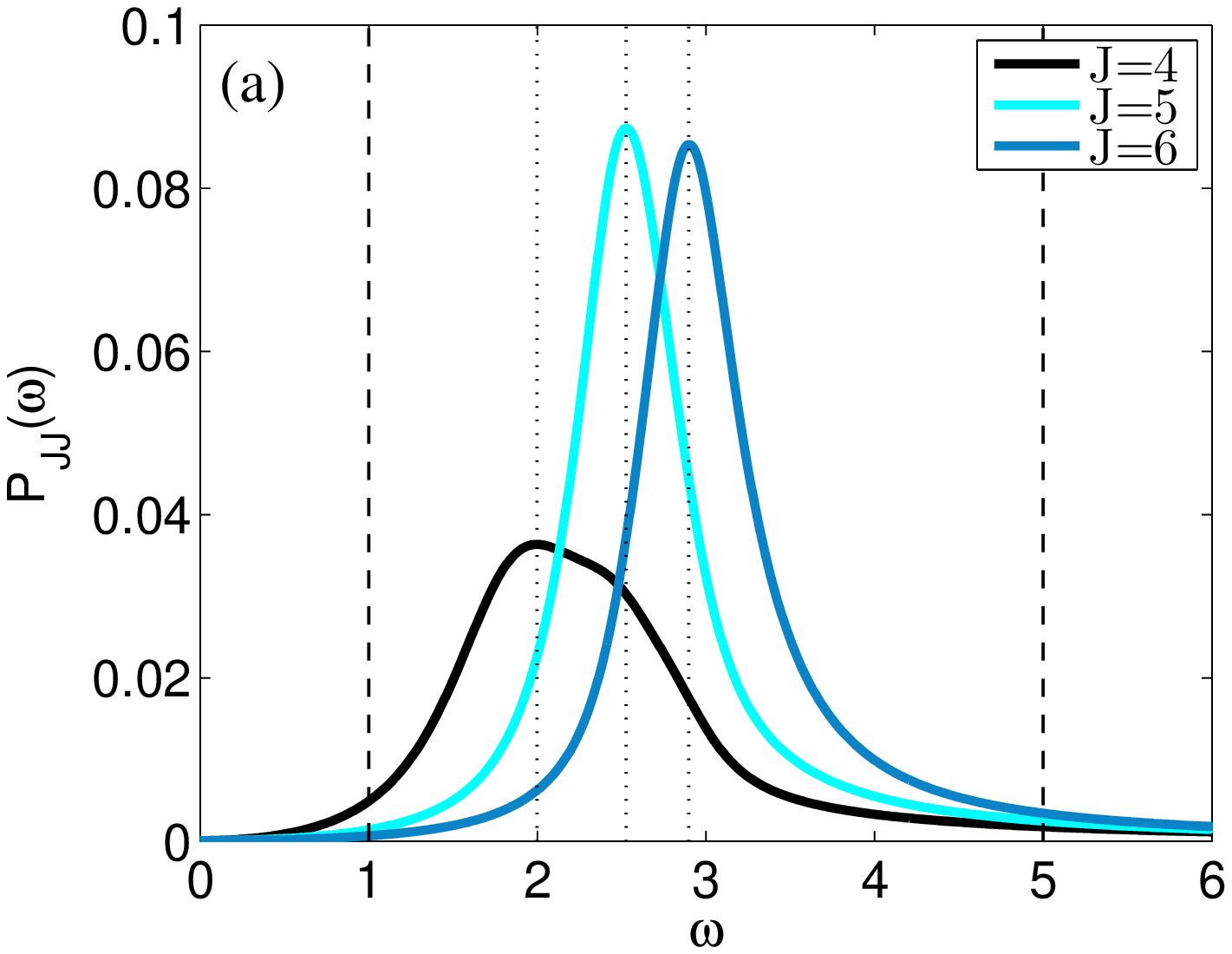}
\includegraphics[width=\columnwidth]{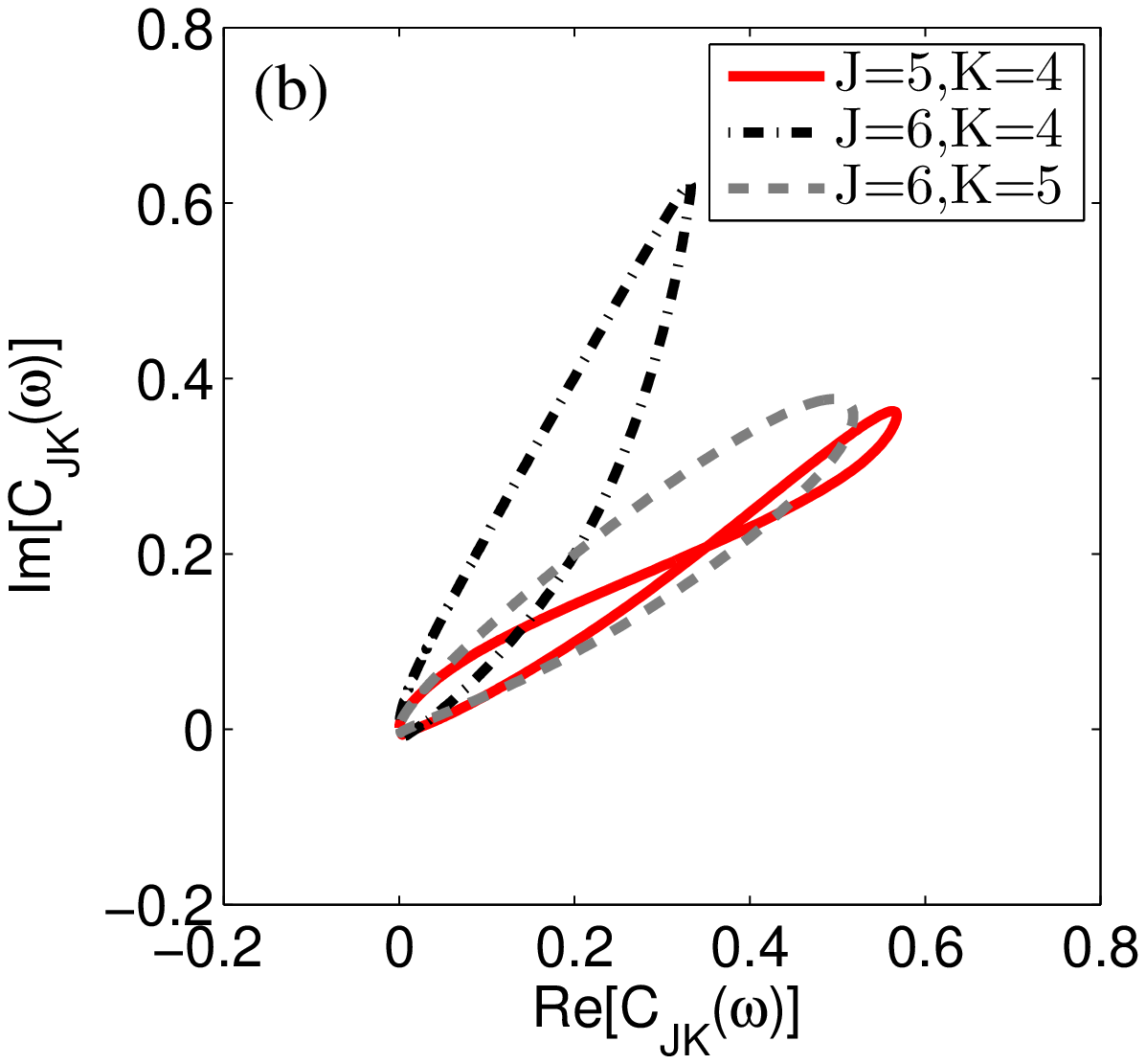}
{\caption{Synchronisation and phase-lag between infected individuals in three 
cities with different infection rates in the three-city SIR model. (a) Power 
spectra of the fluctuations of infectives in city 1, $P_{44}(\omega)$, in city 
2, $P_{55}(\omega)$, and in city 3, $P_{66}(\omega)$, given by Eq.~(\ref{defn}). 
The dotted and the dashed vertical lines have the same meanings as in Fig.~2. 
(b) Parametric plot of the CCFs given by Eq.~\ref{CCF}, $C_{54}(\omega)$, 
$C_{64}(\omega)$, and  $C_{65}(\omega)$, in the complex plane. Parameters: 
$\beta_1=7(\gamma+\mu)$, $\beta_1\colon\beta_2\colon\beta_3=7\colon12\colon17$,
$N_i=10^6$, $\mu=1/50$ 1/y, $\gamma=365/13$ 1/y, and $f_{ij}=0.005$, where $i=1,2,3$ and $i\neq j$.}}
\label{fig5}
\end{figure}

\begin{figure}
\centering
\includegraphics[width=\columnwidth]{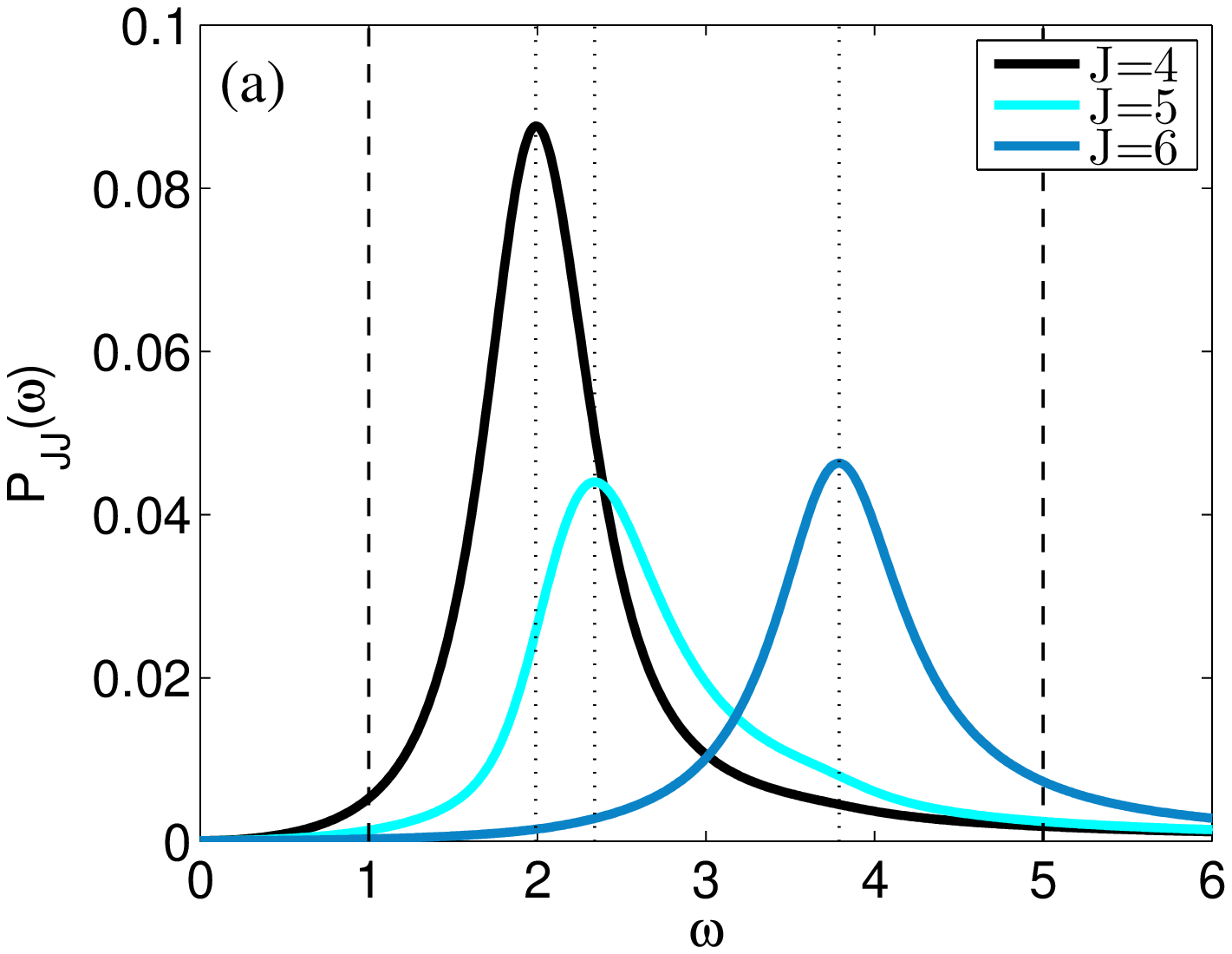}
\includegraphics[width=\columnwidth]{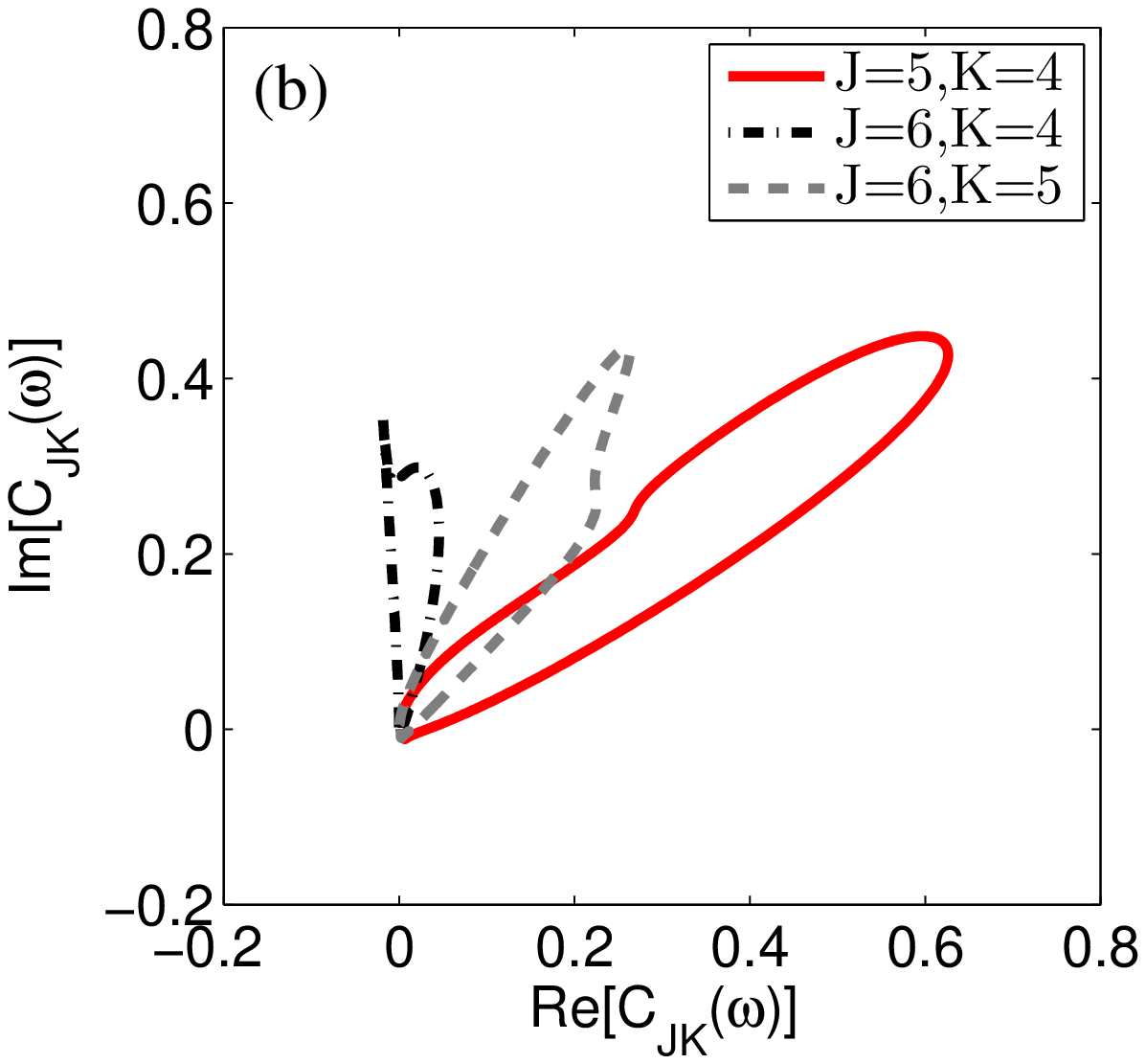}
{\caption{An example of synchronisation and phase-lag between infected 
individuals in three cities in the case when all parameter values are 
different. The description and coloring are as in Fig.~5. Parameters: 
$\beta_1=7(\gamma+\mu)$, $\beta_1\colon\beta_2\colon\beta_3=7\colon12\colon27$,
$N_1=2\times10^6$, $N_1\colon N_2\colon N_3=2\colon 1\colon 0.5$, $\mu=1/50$ 1/y, 
$\gamma=365/13$ 1/y, $f_{12}=f_{31}=0.001$, $f_{13}=f_{23}=0.002$, $f_{21}=0.01$, and $f_{32}=0.005$.}}
\label{fig6}
\end{figure}

The analysis carried out in Sec.~\ref{2cities} can be extended to three or 
more cities. In this Section we will illustrate this by investigating two 
examples for three cities. We have checked that the approximate analytic 
description is again in good agreement with the results of simulations, and 
the plots show the analytic results only, to avoid cluttering. 

In the case of Fig.~5, the population sizes are equal, the demographic coupling
parameters $f_{ij}$ are all equal, but the three transmission rates $\beta _j$,
$j=1,2,3$ are  different. The power spectra of the fluctuations of infectives 
in each city, $P_{JJ}$, $J=4,5,6$, shown in Fig.~5 (a), behave
as in the example of Fig.~2, as do the coherence and the phase spectra. The 
ranges of $C_{54}$, $C_{64}$ and $C_{65}$ in the complex plane are plotted in 
Fig.~5 (b). The phase lags between cities $4$ and $5$ and $5$ 
and $6$ are approximately the same in this case.

In the example of Fig.~6 we have considered different population sizes and
coupling parameters $f_{ij}$ and we have also increased the difference between 
the transmission rates $\beta _j$, $j=1,2,3$ with respect to the previous 
example. The results for the fluctuation power spectra and for the ranges of 
$C_{54}$, $C_{64}$ and $C_{65}$ in the complex plane are plotted in 
Fig.~6 (a) and (b), respectively. In this case we find three different phase lags 
for the correlated fluctuations in the three city pairs.

These phase lag effects have implications for estimates of the duration of an 
epidemic and of the likelihood of disease extinction. In a metapopulation model
with different transmission rates epidemic bursts should last longer, and 
disease extinction should occur less frequently than in a single population 
with the same overall size. 

\section{Conclusions}
\label{final}
In this paper we have considered a stochastic metapopulation version of a
susceptible-infected-recovered model representing infection dynamics in 
different demographically coupled urban centers. We derived the state 
transition rates of the stochastic process from a microscopic model for the 
mobility of individuals between cities. Similar rate equations based on a 
phenomenological coupling parameter have been proposed in the literature. Here 
we have adopted a bottom-up approach, as an illustration of how the parameters 
of the general model can be related with the mobility patterns of the 
sub-populations.

The correlations between the fluctuations in the number of infected in 
different cities were studied by means of the complex coherence function. It 
was found that, if the infection rate differed from city to city and the 
coupling was not too strong, oscillations were observed which were synchronised
with a well defined phase lag between cities. We also showed that for realistic 
population sizes this effect was well described analytically by the linear 
noise approximation.

The combined effect of stochasticity and demographic coupling as a possible
driver of correlations and spatial patterns that are missed by traditional
epidemic models was suggested long ago~\cite{lloyd_may,mech_approach}. This 
idea has been left largely unexplored because these early studies based on 
computer simulations took the infection rate to be the same in different cities.
In this case, as we have seen, the fluctuations are trivially synchronized, 
with no phase lag. However, the infection rate is a phenomenological parameter 
that depends not only on the disease but also on the rate of potentially 
infectious contacts that characterize a given social environment. In 
particular, a strong positive correlation between measles transmission rate and
population density was confirmed in recent study based on daily monitoring of 
urban populations~\cite{measles_satel}. The conditions found in this paper
for a non-trivial phase relation between the disease incidence fluctuations 
in connected urban centres can therefore be considered realistic in many 
settings.

\begin{acknowledgments}
We thank Arkady Pikovsky and Lawrence Sheppard for useful discussions.
Financial support from the Portuguese Foundation for Science and Technology 
(FCT) under Contract No.~POCTI/ISFL/2/261 is gratefully acknowledged. G.R. was 
also supported by FCT under Grant No.~SFRH/BPD/69137/2010.
\end{acknowledgments}

\begin{appendix}

\section{Proof that $P^{(4)}$ is real when the infection rates are equal}
\label{append}
In this Appendix we will show that the matrix $P_{JK}$ defined by 
Eq.~(\ref{defn}) is real for $J,K=n+1,\ldots,2n$ in the case where the 
infection rates for each city are equal: $\beta_{j}=\beta$ for all $j$. This
shows that the cross-spectra for fluctuations in the number of infected 
individuals is real, and so that in this case there is either no phase lag 
or a phase lag of $\pi$. We would expect that the former situation, i.e.~no
phase lag, would hold, and we have explicitly checked this for the case of 
two cities. For general $n$ we have a partial proof that the 
phase lag is zero, and work is underway to complete it. 

The results for the situation where $\beta_{j}=\beta$ for all $j$ can be
inferred from the results of the current paper, and are given explicitly in
Ref.~\cite{epidcities}. They are given in terms of the matrices $A$ and $B$ 
at the fixed point and can be most easily expressed in terms of four 
$n \times n$ submatrices by writing 
\begin{equation}
A =\left[\begin{array}{c|c} 
A^{(1)} & A^{(2)} \\ \hline 
A^{(3)} & A^{(4)} 
 \end{array}\right],
\label{blockA}
\end{equation}
and similarly for $B$ and $\Phi(\omega) \equiv -i\omega I - A$. Here $I$ is the 
$n \times n$ identity matrix. The matrices $\Phi^{(1)}$ and $\Phi^{(3)}$ are
proportional to the identity matrix and we write them as $\rho I$ and 
$\sigma I$ respectively. The inverse of $\Phi(\omega)$ then takes the simple 
form
\begin{equation}
\Phi^{-1}(\omega)  =\left[\begin{array}{c|c} 
-\Phi^{(4)} & \Phi^{(2)} \\ \hline 
\sigma I & - \rho I 
\end{array}\right]
\left[\begin{array}{c|c}
Y & 0 \\ \hline
0 & Y 
\end{array} \right],
\label{block_inverse}
\end{equation}
where $Y^{-1}$ is the $n \times n$ matrix $\sigma \Phi^{(2)} - \rho \Phi^{(4)}$. 

In the calculation of $P_{JK}(\omega)$ the matrices $Y$ appear in the 
combination
\begin{equation}
Z  = \left[\begin{array}{c|c}
Y & 0 \\ \hline
0 & Y 
\end{array} \right]
\left[\begin{array}{c|c} 
B^{(1)} & B^{(2)} \\ \hline 
B^{(3)} & B^{(4)} 
\end{array}\right]
\left[\begin{array}{c|c}
Y^{\dag} & 0 \\ \hline
0 & Y^{\dag} 
\end{array} \right].
\label{blockZ}
\end{equation}
However in the case we are interested in \cite{epidcities}, the submatrices 
$B^{(\alpha)}$, $\alpha=1,\ldots,4$ are diagonal: 
$B^{(\alpha)}_{jk} \equiv B_{\alpha}\delta_{jk}$ with $j,k=1,\ldots,n$. Therefore
\begin{equation}
Z  = \left[\begin{array}{c|c} 
B_{1} Y Y^{\dag} & B_{2} Y Y^{\dag} \\ \hline 
B_{3} Y Y^{\dag} & B_{4} Y Y^{\dag} 
\end{array}\right].
\label{blockZZ}
\end{equation}

The quantity we wish to study, $P_{JK}(\omega)$, is now given by
\begin{equation}
P  = \left[\begin{array}{c|c}
-\Phi^{(4)} & \Phi^{(2)} \\ \hline
\sigma I & - \rho I
\end{array} \right]
\left[\begin{array}{c|c} 
Z^{(1)} & Z^{(2)} \\ \hline 
Z^{(3)} & Z^{(4)} 
\end{array}\right]
\left[\begin{array}{c|c}
- \Phi^{(4)\dag} & \bar{\sigma} I \\ \hline
\Phi^{(2)\dag} & - \bar{\rho}I
\end{array} \right],
\label{blockP}
\end{equation}
where $Z^{\alpha}$ is one of the four submatrices given in Eq.~(\ref{blockZZ}).

It follows from Eq,~(\ref{blockP}) that
\begin{equation}
P^{(4)} = \sigma^2 Z^{(1)} - \sigma \left( \rho + \bar{\rho} \right) Z^{(2)} + 
| \rho |^{2} Z^{(4)},
\label{P_4}
\end{equation}
where we have used $Z^{(2)}=Z^{(3)}$ which follows from the the symmetry of the 
matrix $B$ ($B_2 = B_3$), and also used the reality of $\sigma$. We now show 
that the matrix $Z$ is real, and so that $P^{(4)}$ is real.

To do this we write down the explicit forms for the entries of $\Phi(\omega)$. 
They are~\cite{epidcities}
\begin{eqnarray}
\rho &=& \frac{\mu \beta}{\mu+\gamma} -i\omega , \nonumber\\  
\sigma &=& \mu -  \frac{\mu \beta}{\mu+\gamma},\nonumber \\
\Phi^{(2)} &=& \left( \mu +\gamma \right) \chi, \nonumber\\ 
\Phi^{(4)} &=& \left( \mu + \gamma - i\omega \right)I - 
\left( \mu +\gamma \right) \chi,
\label{explicit_Phi}
\end{eqnarray}
where $\chi$ is an $n \times n$ matrix given by 
$\chi_{jk} = \left( N_j/N_k \right)^{1/2} c_{jk}$, and is real and symmetric
(using Eq.~(B5) of \cite{epidcities}). Therefore from the definition of the
matrix $Y$
\begin{equation}
Y^{-1} = G(\omega)I + H(\omega) \chi,
\label{inverseY}
\end{equation}
with 
\begin{equation}
G(\omega) = - \left( \frac{\mu\beta}{\mu+\gamma} -i\omega\right)
\left(\mu + \gamma - i\omega\right)
\end{equation} and
\begin{equation}
H(\omega) = \left(\mu+\gamma\right) \left(\mu - i \omega\right).
\end{equation}
\noindent{}Using the symmetry and reality of
the matrix $\chi$,
\begin{eqnarray}
Y^{-1 \dag} Y^{-1} &=& |G(\omega)|^{2}I + |H(\omega)|^{2}\chi^{2} \nonumber \\
&+& \left\{ G(\omega) \overline{H} (\omega) + \overline{G} (\omega) H(\omega) \right\}\chi , 
\label{product_Y}
\end{eqnarray}
which is clearly real. Therefore the inverse of this matrix, $Y Y^{\dag}$ is
also real. From Eq.~(\ref{blockZZ}) it follows that $Z$ is real, since $B$ 
is real.

We end by showing that the cross-spectra for the fluctuations in the number 
of susceptible individuals is also real in this case. Although we have not
explicitly investigated this quantity in the main text, it is also of 
interest, and by analogy with $P^{(4)}$ we would expect it to be real too.
From Eq,~(\ref{blockP})
\begin{eqnarray}
P^{(1)} &=& \Phi^{(4)}Z^{(1)}\Phi^{(4)\dag} - \Phi^{(4)}Z^{(2)}\Phi^{(2)\dag}
\nonumber \\
&-& \Phi^{(2)}Z^{(3)}\Phi^{(4)\dag} + \Phi^{(2)}Z^{(4)}\Phi^{(2)\dag}.
\label{P_1}
\end{eqnarray}
Since from Eq.~(\ref{explicit_Phi}), 
$\Phi^{(4)} = ( \mu + \gamma -i\omega ) I - \Phi^{(2)}$ and $\Phi^{(2)}$ is real,
and since the $Z^{\alpha}$ are real, the imaginary part of $P^{(1)}$ is given
by 
\begin{eqnarray}
{\rm Im}P^{(1)} &=& \omega \left\{ Z^{(1)}\Phi^{(2)} - \Phi^{(2)}Z^{(1)} \right.
\nonumber \\
&+& \left. Z^{(2)}\Phi^{(2)} - \Phi^{(2)}Z^{(3)} \right\}.
\label{ImP_1}
\end{eqnarray}
Writing $Z^{\alpha}=B_{\alpha} YY^{\dag}$ and $\Phi^{(2)} = (\mu + \gamma)\chi$ 
yields
\begin{equation}
{\rm Im}P^{(1)} = \omega (\mu + \gamma) \left( B_1 + B_2 \right)
\left\{ \left( YY^{\dag} \right) \chi - \chi \left( YY^{\dag} \right) \right\},
\label{ImagP_1}
\end{equation}
where we have used $B_{3}=B_{2}$. Now from Eq.~(\ref{product_Y}), the matrix
$Y^{-1 \dag} Y^{-1}$ is a sum of matrices proportional to the identity, $\chi$
and $\chi^2$. Therefore it commutes with the matrix $\chi$, that is,
$\chi Y^{-1 \dag} Y^{-1} = Y^{-1 \dag} Y^{-1} \chi$, from which it follows that
$YY^{\dag} \chi = \chi  YY^{\dag}$, and so from Eq.~(\ref{ImagP_1}), 
${\rm Im}P^{(1)}=0$.

\end{appendix}

%\bibliography{SIR}

\end{document}